# Big Data based Research on Mechanisms of Sharing Economy Restructuring the World

Dingju Zhu

South China Normal University, Guangzhou, China
Corresponding Author's Email:  zhudingju@m.scnu.edu.cn

**Abstract**

Many researches have discussed the phenomenon and definition of sharing economy, but an understanding of sharing economy's reconstructions of the world remains elusive. We illustrate the mechanism of sharing economy's reconstructions of the world in detail based on big data including the mechanism of sharing economy's reconstructions of society, time and space, users, industry, and self-reconstruction in the future, which is very important for society to make full use of the reconstruction opportunity to upgrade our world through sharing economy. On the one hand, we established the mechanisms for sharing economy rebuilding society, industry, space-time, and users through qualitative analyses, and on the other hand, we demonstrated the rationality of the mechanisms through quantitative analyses of big data.

**Keywords**

Sharing economy；Reconstruct；Sustainability；The World

## Introduction

The human economy is constantly evolving, but it is constrained by limited natural resources and energy supplies. It is difficult for human needs to meet unrestricted satisfaction. Sharing economy is a panacea for resolving this dilemma. It makes people's needs by increasing resource utilization. The sharing economy not only improves resource utilization, but also improves service quality and user experience. The sharing economy is reconstructing the world.

The reason why the sharing economy started to rise from sharing bikes in China is that bicycles were imported from abroad during the Qing Dynasty, and became popular after the founding of New China. Therefore, Chinese people are very familiar with bicycles and have a feeling of hardship. With the rapid development of China's economy, private cars have gradually replaced bicycles, but traffic jams and parking difficulties have made bicycles once more important because bicycles help ease traffic pressure. Although bicycles also have the problem of inconvenience of parking, the appearance of shared bicycles makes it possible for shared bicycles can be parked everywhere and bring great convenience to people. However, the parking of shared bicycles has also caused problems affecting the city appearance and occupying lanes and sidewalks. On the one hand, the government is required to reconstruct the urban transport facilities. On the other hand, citizens are required the reconfiguration of their bicycle travel habits. The citizens are required to observe the parking rules when using shared bicycles and consciously park the shared bicycles.

Unexpectedly, the shared bicycles that have become popular in China in recent years have not only evoked the enthusiasm of the Chinese people for cycling, but also evoked the return of enthusiasm for cycling by people of other countries, and even triggered an upsurge of sharing everything.

The sharing economy lacks a common definition[1]. Sharing something with others is a long-standing practice for humankind [2]. Sharing is a phenomenon that is as old as humanity, and collaborative consumption and the "sharing economy" are phenomena born in the age of the Internet [3]. Although individuals traditionally often regard ownership as the ideal way to obtain products, more and more consumers temporarily pay to access shared services of products rather than purchase or own products[4]. The development of the sharing economy at the beginning of the 21st century is based on the revolution of the Internet by connecting consumers and unused resources via the Internet[5]. Participants can share, rent, lend, trade, service, and transport through the Internet[6]. Information and communication technology has led to collaborative consumption[7]. The sharing service platform can achieve point-to-point access to shared products and services based on information technology [8]. The sharing economy is causing dramatic changes in the retail and service structure [9]. Share-economy companies are destroying traditional industries around the world [10]. At the same time, the rise of the sharing economy has created many new industry competitions such as Airbnb, Uber, Lyft and Sidecar [11]. "Sharing economy" platforms such as Airbnb have recently thrived in tourism[12]. The sharing economy is a potential new way of sustainable development[13].

Many studies have discussed the phenomenon and definition of sharing economy. The sharing economy is an economic form of service innovation [14-17] to achieve the sharing of resources among consumers[18]. The sharing economy has spread to various fields such as bicycles and automobiles[19]. Different scholars have different views on the future form and prospect of the sharing economy[20]. Researchers are struggling to find a model that can describe the sharing economy[21,22].

However, existing studies on sharing economy's reconstructions of society, time and space, users, industry, and sharing economy's reconstruction in the future are still blank, but the sharing economy's reconstruction on the world which we found is very important for society to make full use of the reconstruction opportunity to upgrade our world trough sharing economy.

## Methods

We studied the model of the sharing economy to rebuild the world through model building and big data analysis. Our data comes from statistics and analysis of big data on the Internet. Internet big data is generated by people all over the world, so it can reflect social dynamics and people's participation in it.

We first use the modeling method to build mechanisms for the sharing economy to restructure society, enterprises, and users. We then counted relevant data from the Internet and performed analysis to support the rationality of our mechanisms. Our data sources come from official statistics on the one hand, such as population and GDP, on the other hand, statistics from related information on the Internet. Both data are very important for verifying our mechanisms.

When demonstrating the social reconstruction mechanism of the sharing economy, we conducted data statistics through the Internet and combined the official data to analyze the correlation between the state of the sharing economy in various countries and social data in various aspects in each country, and then the internal mechanism of the correlation is analyzed and explained. The sharing economy's mechanism for rebuilding society shows that GDP and population are most important social foundations for the existence and development of the sharing economy. When demonstrating the mechanism of reshaping the space-time of the sharing economy, we prove the complementary mechanism of resources and consumption in time and space under the sharing economy model by sharing bicycle data in a certain period of time and in a certain region. When demonstrating the reconstruction of user consumption by the sharing economy, we also analyzed the sharing mechanism between a large number of users and bicycle resources through the data of shared bicycles over a period of time and region, and pointed out the current deficiencies of the current sharing economy industry. In analyzing the reconstruction of the sharing economy to the industries, we analyzed and evaluated the current industrial structure of the sharing economy through the amount of Internet information about shared products, shared platforms, and shared facilities. In the analysis of the self-restructuring mechanism of the sharing economy, we give specific examples to illustrate the two trends of the sharing economy in the future.

# Results

**Mechanism of sharing economy's reconstruction of society**

The sharing economy allows users to access any service in a shared manner anytime, anywhere, just as we use tap water and electricity in our daily lives. Open the mobile phone app anytime and anywhere to get the sharing services you want, including sharing cars, sharing bicycles, sharing KTV, sharing basketball. Users do not need to buy products, install products, and do not need to maintain and upgrade products. Only need to apply for sharing at the time of use, return when not in use. In the rush season, sharing service platforms are not short of resources. This is because the resources of the sharing service platform are shared, so that resources in different time and space can be complementary. As a result, shared resources are inexhaustible for users. In the off-season, there is no free resource, a user or a platform's free resources will be timely scheduled to other users or other platforms to make full use of. It is precisely because the sharing economy has such a service sharing and scheduling mechanism that the sharing economy has higher resource utilization than the traditional economy. In the sharing economy, the same amount of resources can be used by more users. In turn, the same users only need to use fewer resources, so as to achieve the purpose of saving resources.

The wave of the sharing economy has swept China in recent years and is also sweeping the world. The reason is that the conditions for the sharing economy have matured. Users can apply for sharing services as long as they can connect to the Internet. According to the data from the Ministry of Communications of the People's Republic of China, as of July 2017, there are nearly 70 bicycle sharing companies in China, and more than 16 million vehicles have been deployed in more than 100 cities across the country, with more than 130 million registered users and cumulative services exceeding 1.5 billion. According to the "China's Sharing Economy Development Annual Report" released on the February 28th,2019 by the China National Information Center Sharing Economy Research Center, the share economy market transaction value in 2018 was 294.2 billion RMB, an increase of 41.6% over the previous year; the number of platform employees was 5.98 million , an increase of 7.5% over the previous year; the number of participants in the sharing economy was about 760 million, of which the number of service providers was about 75 million, a year-on-year increase of 7.1%. Most industries in the sharing economy in 2019 have begun to move from large-scale horse race enclosures to refined operations. The most important driving force of the sharing economy is to increase resource utilization and energy efficiency, because human beings have realized that resources and energy on earth are not inexhaustible, so they need to save resources and energy, and sharing service platforms can effectively avoid repeated construction and idleness of resources. At the same time, the increasingly diversified and personalized needs of users cannot be met under the traditional economy model, because the limited resources in mutually independent platforms and services are difficult to cope with diverse and constantly changing user needs. Only through the rapid deployment of massive resources in the sharing service platform can we quickly meet the personalized needs of a large number of users.

The current sharing economy has become an important technological revolution and business model that is highly valued by governments and companies of all countries. The sharing economy is a new thing. The sharing economy in each country is just starting. Therefore, countries with different degrees of development can stand on the same starting line for the transformation and upgrading of the traditional economy to the sharing economy. The success of the countries represented by China in the sharing economy has made the sharing economy attracting attention from all countries, especially those leaders who are concerned about resource shortages and energy crisis, and have no capital to purchase products. Companies have excess resources during their off-season, and companies that lack resources during their busy season, either view the sharing economy as a life-saving straw or view the sharing economy as an unprecedented opportunity for development. The "China Sharing economy Development Report 2017" released by China's State Information Center shows that the rapid growth of China's sharing economy has made important contributions to economic development, innovation, and employment expansion. The report points out that in the next few years, the sharing economy will still maintain 40% annual growth rate. By 2020, the share of economic transactions will account for more than 10% of GDP and will increase to around 20% by 2025.

The sharing economy can solve a series of problems in the traditional economy and can inject new momentum into the development of traditional industries. The main reasons are as follows: First, the current accumulation of resources has reached a massive level. This is the basis of sharing economy for no matter how skillful the sharing technique is, if there are not enough resources for sharing, it is also impossible to provide enough shared services to users. Second, the accumulation of information technologies in Internet+, big data, and artificial intelligence has made it possible to share resources. Third, the current economy development has been increasingly affected by resource shortages and energy crises, and the more efficient use of resources through sharing is an effective way to solve this development dilemma. Fourth, the sharing economy can be flexibly allocated to users by integrating platform resources to increase resource utilization, reduce idleness and waste of resources, and at the same time quickly meet the personalized needs of various users, which cannot be achieved by the traditional economy models.

We chose some typical countries，and sort the amount of information and negative information related to the sharing economy, GDP, population, and land area of these countries, as shown in table 1 and 2. The amount of information related to the sharing economy can reflect the country's enthusiasm and level for the development of the sharing economy. The amount of negative information related to the sharing economy can reflect this country's policies and people's resistance to the sharing economy. Because the amount of negative information is a hindrance to the development of the sharing economy, we have used this indicator in reverse order. Because the information on the Internet is increasing every day, we have count the information that is publicly available online until January 11, 2020. The amount of the related informations and the negative informations are counted from the informations in Internet. The newest GDP data is the statistics data of 2018 which we list in our table, and the population and the area shown in our table are the newest data in January 1, 2020.

**Table 1: The data related to the sharing economy of some typical countries**

| Country | Related Informations | GDP | Population | Area (km2) | Negative Informations |
|---|---|---|---|---|---|
| China | 139,000,000 | 13,457,267 | 1,408,526,449 | 9,596,960 | 3,630,000 |
| America | 192,000,000 | 20,513,913 | 332,865,306 | 9,833,520 | 4,990,000 |
| Japan | 118,000,000 | 5,070,626 | 125,938,348 | 377,835 | 18,700,000 |
| Germany | 115,000,000 | 4,029,140 | 81,453,631 | 357,021 | 2,520,000 |
| France | 108,000,000 | 2,794,696 | 65,569,000 | 547,030 | 4,870,000 |
| Russia | 82,100,000 | 1,576,488 | 146,570,133 | 17,098,242 | 12,700,000 |
| Canada | 169,000,000 | 1,733,706 | 37,281,000 | 9,984,670 | 3,090,000 |
| Australia | 153,000,000 | 1,427,767 | 25,220,000 | 7,686,850 | 2,910,000 |
| India | 191,000,000 | 2,689,992 | 1,387,297,452 | 3,287,590 | 3,210,000 |
| Bengal | 29,800,000 | 286,275 | 169,872,008 | 144,000 | 1,630,000 |
| Egypt | 92,400,000 | 249,471 | 108,071,377 | 1,001,450 | 9,660,000 |
| Sudan | 37,000,000 | 33,249 | 43,375,000 | 1,861,484 | 3,130,000 |
| Israel | 30,200,000 | 350,851 | 8,448,300 | 20,770 | 7,380,000 |
| Venezuela | 55,800,000 | 371,337 | 32,691,000 | 912,050 | 4,620,000 |
| Poland | 31,900,000 | 544,959 | 38,646,000 | 312,685 | 7,190,000 |
| Mongolia | 31,400,000 | 12,724 | 3,200,200 | 1,564,116 | 5,770,000 |
| Korea | 121,000,000 | 1,530,751 | 51,285,000 | 99,600 | 11,100,000 |
| United Kingdom | 57,900,000 | 2,808,899 | 66,366,000 | 244,820 | 15,800,000 |

**Table 2: The sorted rank related to the sharing economy of some typical countries**

| Country | Related Informations | GDP | Population | Area km2 | Negative Informations |
|---|---|---|---|---|---|
| China | 5 | 2 | 1 | 4 | 7 |
| America | 1 | 1 | 3 | 3 | 10 |
| Japan | 7 | 3 | 6 | 12 | 18 |
| Germany | 8 | 4 | 8 | 13 | 2 |
| France | 9 | 6 | 10 | 11 | 9 |
| Russia | 11 | 9 | 5 | 1 | 16 |
| Canada | 3 | 8 | 14 | 2 | 4 |
| Australia | 4 | 11 | 16 | 5 | 3 |
| India | 2 | 7 | 2 | 6 | 6 |
| Bengal | 18 | 15 | 4 | 16 | 1 |
| Egypt | 10 | 16 | 7 | 9 | 14 |
| Sudan | 14 | 17 | 12 | 7 | 5 |
| Israel | 17 | 14 | 17 | 18 | 13 |
| Venezuela | 13 | 13 | 15 | 10 | 8 |
| Poland | 15 | 12 | 13 | 14 | 12 |
| Mongolia | 16 | 18 | 18 | 8 | 11 |
| Korea | 6 | 10 | 11 | 17 | 15 |
| United Kingdom | 12 | 5 | 9 | 15 | 17 |

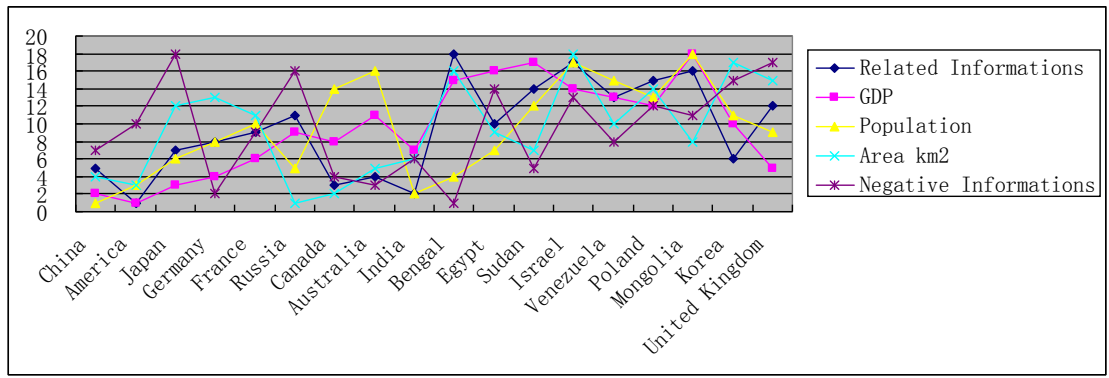

**Figure 1. The sorted rank related to the sharing economy of some typical countries**

As shown in figure 1, we found that the amount of information related to the sharing economy is positive correlation with the GDP in most cases(13/17). Because the more wealth and assets a country has, the more assets and services it can use to share. For example, American GDP is larger than Chinese GDP and thus American informations are more than Chinese informations, and American GDP is larger than Japanese GDP and thus American informations are more than Japanese informations.

From figure 1, we can also found that the amount of information related to the sharing economy is positive correlation with the population in most cases(11/17). Because the more people there are in a country, and the needs between people are overlapped, these overlapped needs are the basis for sharing, then the greater the sharing among different people in this country. For example, American population is larger than Japanese population and thus American informations are more than Japanese informations, and Japanese population is larger than German population and thus Japanese informations are more than German informations.

From figure 1, we can also found that the amount of information related to the sharing economy is positive correlation with the area in slightly more cases(9/17). Because private items such as bicycles are difficult to carry too far, the larger the area, the greater the need for sharing services when traveling. For example, American area is larger than Japanese area and thus American informations are more than Japanese informations, and Japanese area is larger than German area and thus Japanese informations are more than German informations.

From figure 1, we can also found that the amount of information related to the sharing economy is negative correlation with the amount of negative information in almost half cases(8/17). Because negative information is part of related information, when there is more relevant information, there will naturally be more negative information, but sometimes negative information can play an important role in hindering the development of the sharing economy, thereby inhibiting the generation of related information. For example, Korean GDP, population and area are all less than British GDP, population and area, however Korean informations are more than British informations, which is caused by Korean negative informations are less than British negative informations, and from the table we can see that the British negative informations is ranked secondly. This means that the British and the government are more resistant to the sharing economy model, which in turn hinders the development of the sharing economy, and thus cannot take advantage of its GPU, population and area.

**Mechanism of sharing economy's reconstruction of time and space**

Since the shared services platform can utilize the complementarity of different temporal and spatial resources of different users to balance and coordinate the resource sharing among different users, it is possible to increase resource utilization and user satisfaction without adding new resources. In the non-sharing economy mode, because the resources of different users are all owned by themselves, some user resources are insufficient and some user resources are idle in different time periods. This kind of resource imbalance occurs for many reasons, such as the impact of busy season and off-season of services, shortage and surplus of funds, and so on. Sharing the economic model can change this situation. Through the sharing of resources among different users, idle resources of some users can be scheduled to resource-intensive users through a sharing mechanism. In this way, through the mutual sharing of users, mutual assistance and mutual benefit, it not only solves the problem of idle resources for some users, but also improves resource utilization, solves the problem of shortage of resources of other users, and satisfies the needs of users. When the number of users is large, the probability that different users' scarce resources and idle resources occur in different time and space should be similar, so that the peak and low periods of resource utilization in different time and space can cancel each other out, so as to achieve the balance of resources. Through the sharing of resource services, idle resources can no longer be idle, and scarce resources are no longer in short supply.

The sharing economy can use the complementarity of temporal differences for the services corresponding to the resources in time are shared among users, while the traditional economy can not use the complementarity of temporal differences for the resources in time are belong to different users (Figure 2). In the sharing economy model, resource-constrained users do not need to purchase more resources, as long as they pay for shared services during their use, which can save a lot of resource acquisition costs. Conversely, in the traditional economy, if users purchase more resources during the peak season to ease the pressure of resource constraints, they will not only consume a lot of funds to purchase resources, but these newly purchased resources will inevitably idle during the off-season and cause waste of resources. At the same time, in the sharing economy model, users with idle resources can provide idle resources to users who are in short supply by sharing schedules. Users who use shared resources need to pay a certain shared service fee to users who provide shared resources, so that idle resources are fully utilized and resources utilization are increased. Users can gain additional revenue by sharing idle resources. The flow of these resources and the sharing of services are not necessarily carried out directly between users, but are generally shared and organized through shared services platforms and their operators.

Users with limited resources do not need to spend more money to purchase resources. They only need to apply for services corresponding to the resources from the shared services platforms. These services will stop when they are not needed without incurring additional costs. Users with idle resources need not worry about that resources idleness can result in wasted resources and economic losses because users can rent idle resources to the shared services platform which can turn resources into shared services. The shared services platform will return some of the benefits obtained by the shared service to users who provide idle resources.

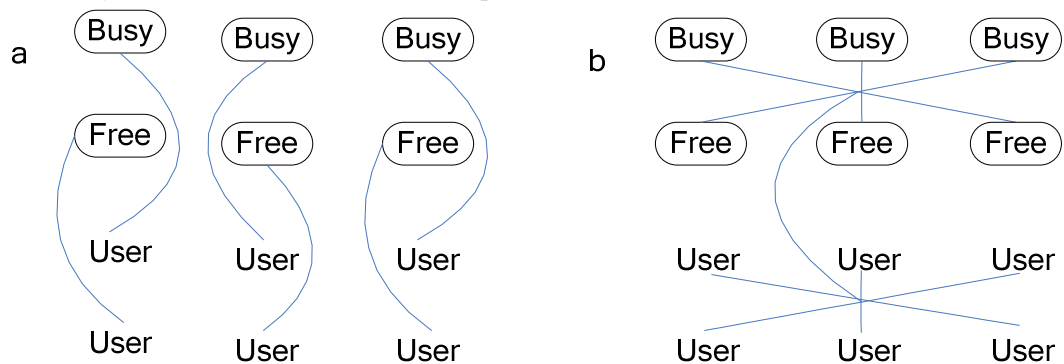

**Figure 2 Temporal differences of different users. (a) traditional economy can not use the complementarity of temporal differences for the resources in time are belong to different users, (b) sharing economy can use the complementarity of temporal differences for the services corresponding to the resources in time are shared among users.**

We selected the bike sharing data of Mobike in Shanghai on August 20, 2017. In the data, the start and end times of the ride are both within August 20, 2017. In order to study the complementarity between free and busy of different users, we selected the user id column, the start time column and the end time column from the data. There are a total of 2286 pieces of data in the bike sharing data of Mobike in Shanghai on August 20, 2017, we only show a few of them as example data shown in Table 3.

**Table 3: Sample data of start time and end time in the bike sharing data of Mobike in Shanghai on August 20, 2017**

| userid | start_time | end_time |
|---|---|---|
| 10080 | 2016-8-20 6:57 | 2016-8-20 7:04 |
| 13745 | 2016-8-20 14:25 | 2016-8-20 14:31 |
| 5564 | 2016-8-20 17:44 | 2016-8-20 18:35 |
| 10915 | 2016-8-20 10:55 | 2016-8-20 11:01 |
| 11510 | 2016-8-20 15:10 | 2016-8-20 15:15 |
| 5198 | 2016-8-20 19:59 | 2016-8-20 20:25 |
| 231 | 2016-8-20 20:48 | 2016-8-20 21:23 |
| 10127 | 2016-8-20 15:16 | 2016-8-20 15:22 |
| 9571 | 2016-8-20 21:09 | 2016-8-20 21:19 |
| 5872 | 2016-8-20 23:40 | 2016-8-20 23:57 |
| 15094 | 2016-8-20 22:18 | 2016-8-20 22:40 |
| 13210 | 2016-8-20 9:08 | 2016-8-20 9:25 |
| 9204 | 2016-8-20 17:18 | 2016-8-20 17:53 |
| 6961 | 2016-8-20 17:18 | 2016-8-20 18:32 |
| 7110 | 2016-8-20 15:45 | 2016-8-20 15:57 |

It can be seen from the table 3 that different users have different starting and ending times. The departure time is the time when the user starts to be busy, and the end time is the time when the user starts to be free. We counted how many users started to be busy and how many users became idle in different time periods, as shown in Table 4 and Figure 3 and Figure 4. The vertical axis of the graph is the number of users, and the horizontal axis of the graph is the time period.

**Table 4: Count how many users started to be busy and how many users became free**

| Time periods | Start to be busy | Start to be free |
| --- | --- | --- |
| 00:00-01:00 | 30 | 23 |
| 01:00-02:00 | 19 | 26 |
| 02:00-03:00 | 15 | 13 |
| 03:00-04:00 | 2 | 4 |
| 04:00-05:00 | 7 | 5 |
| 05:00-06:00 | 14 | 11 |
| 06:00-07:00 | 39 | 31 |
| 07:00-08:00 | 67 | 67 |
| 08:00-09:00 | 103 | 101 |
| 09:00-10:00 | 116 | 103 |
| 10:00-11:00 | 113 | 114 |
| 11:00-12:00 | 103 | 111 |
| 12:00-13:00 | 89 | 88 |
| 13:00-14:00 | 108 | 118 |
| 14:00-15:00 | 113 | 95 |
| 15:00-16:00 | 137 | 123 |
| 16:00-17:00 | 155 | 143 |
| 17:00-18:00 | 203 | 194 |
| 18:00-19:00 | 180 | 179 |
| 19:00-20:00 | 221 | 195 |
| 20:00-21:00 | 155 | 182 |
| 21:00-22:00 | 153 | 160 |
| 22:00-23:00 | 101 | 124 |
| 23:00-24:00 | 43 | 76 |

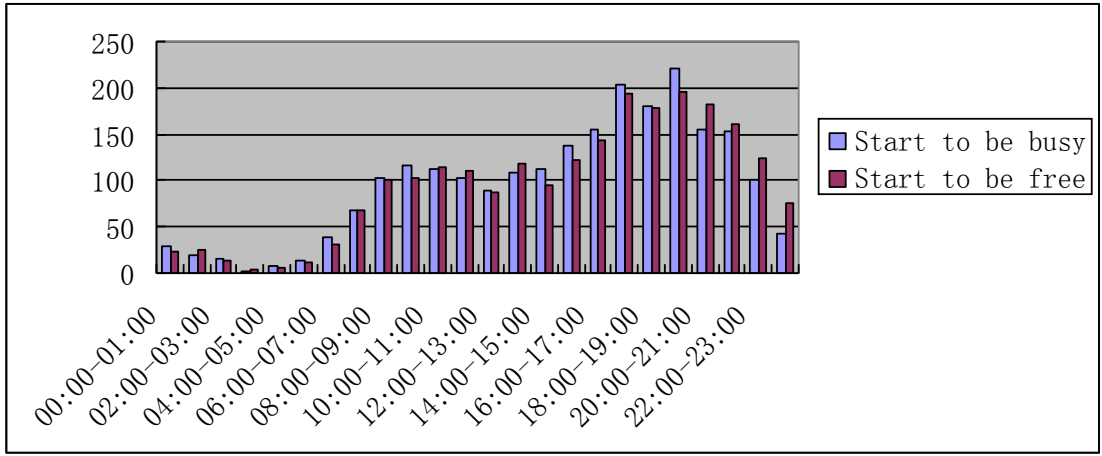

**Figure 3 The histogram of the complementarity between free and busy of different users**

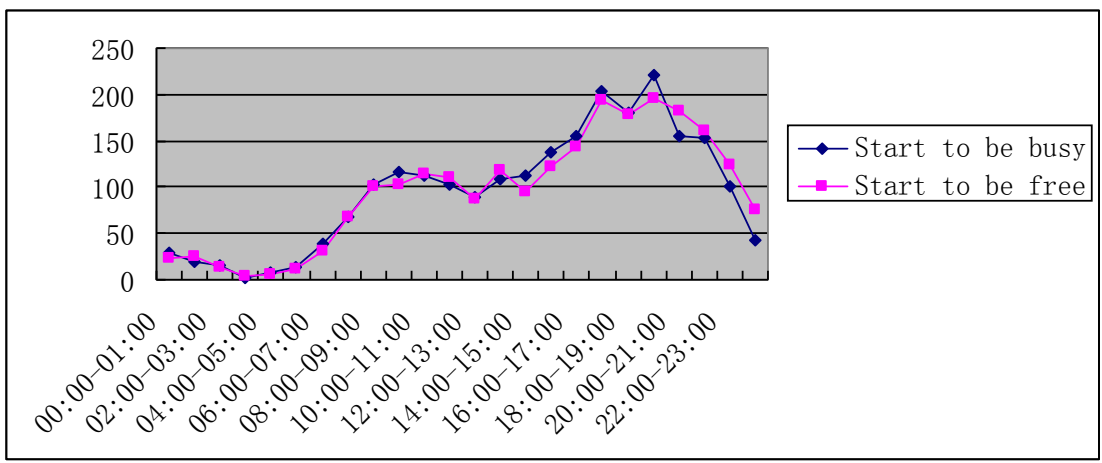

**Figure 4 The chart of the complementarity between free and busy of different users**

  It can be seen from the figure 3,4 that the busy time and the free time of different users can just complement each other. In this way, the resources of the idle users can be released for the busy users to use. The statistical analysis results of this data demonstrate the correctness of the theory we have previously proposed.

  The sharing economy can use the complementarity of spatial differences for the services corresponding to the resources in space are shared among users, while traditional economy can not use the complementarity of spatial differences for the resources in space are belong to different users (Figure 5). The shared services platform can mobilize resources at any time to provide sharing services for specific users. After the service is completed, the resources can be immediately shared with other users to provide services. When the user does not need the service, the service resource is immediately recovered by the shared platform, so the user only needs to pay for the service during use, without having to pay for resources maintenance, service upgrade, etc. that have to be paid in the traditional economic mode. For example, where a shared bike rides, you don't want to ride it, put it there, and you don't need to return it, or stop at a specific parking space. The shared bicycle placed on A can be used by the user to ride from A to B and placed on B. The shared bicycle placed on B can be used by the user to ride from B to A and placed on A, thus forming a complementary difference sets. The form of bicycle sharing has always been operated in some countries as "Pile", such as Citi Bikes. However, "Pile" bicycles can only be placed and recycled at specific locations, and cannot meet the individual needs of all users at the same time. At the same time, the user needs to go to the "Pile" bicycle delivery point to pick up and return the bicycle, causing inconvenience to the user. The Chinese subtly upgraded the "Pile" shared bicycles to " Pileless" shared bicycles, allowing users to ride and return them anywhere. "Pileless" shared bicycles cleverly use the user's differences to achieve the convenience and efficiency of sharing. The success of the " Pileless" bicycle sharing has led many countries to follow "Pileless" shared bicycles. For example, people of insight in the United States call on the United States to share bicycles into the "2.0 era" of "Pileless" development.

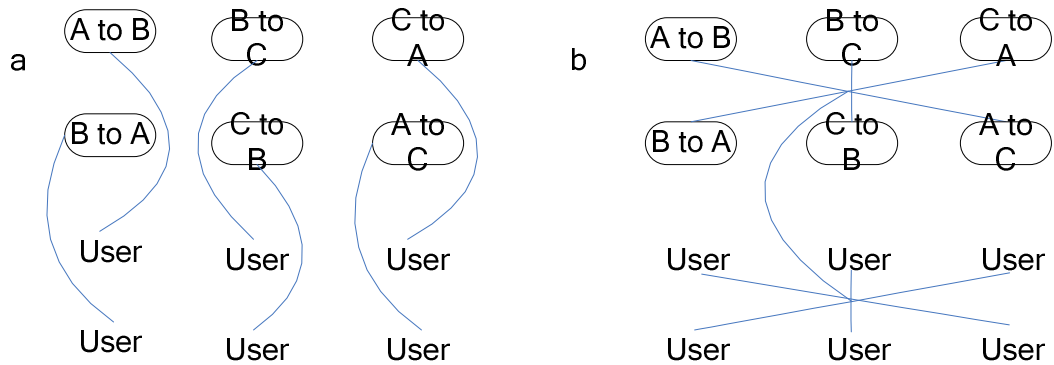

**Figure 5 Spatial differences of different users. (a) traditional economy can not use the complementarity of spatial differences for the resources in space are belong to different users, (b) sharing economy can use the complementarity of spatial differences for the services corresponding to the resources in space are shared among users.**

We selected the bike sharing data of Mobike in Shanghai on August 20, 2017. In the data, the start and end times of the ride are both within August 20, 2017. In order to study the complementarity between start locations and end locations of different users, we selected the user id column, the start location column and the end location column from the data. There are a total of 2286 pieces of data in the bike sharing data of Mobike in Shanghai on August 20, 2017, we only show a few of them as example data shown in Table 5. Because the longitude and latitude of Tianjin area are near (112,31). Therefore, in order to show the characteristics of the data more clearly when drawing, we removed the value before the decimal point, but took the value after the decimal point to make the scatter plot, as shown in table 6 and fiture 6-8. The data of different start locations of different users are shown in FIGURE 6; the data of different destinations of different users are shown in FIGURE 7; the data of different start and end locations of different users are superimposed and shown in FIGURE 8.

**Table 5: Sample data of start location and end location in the bike sharing data of Mobike in Shanghai on August 20, 2017**

| userid | start_location_x | start_location_y | end_location_x | end_location_y |
|---|---|---|---|---|
| 10080 | 121.348 | 31.389 | 121.357 | 31.388 |
| 13745 | 121.432 | 31.18 | 121.427 | 31.184 |
| 5564 | 121.477 | 31.267 | 121.441 | 31.304 |
| 10915 | 121.447 | 31.319 | 121.444 | 31.314 |
| 11510 | 121.429 | 31.349 | 121.422 | 31.35 |
| 5198 | 121.431 | 31.277 | 121.443 | 31.281 |
| 231 | 121.402 | 31.263 | 121.351 | 31.286 |
| 10127 | 121.375 | 31.247 | 121.376 | 31.25 |
| 9571 | 121.491 | 31.229 | 121.48 | 31.224 |
| 5872 | 121.443 | 31.262 | 121.422 | 31.266 |
| 15094 | 121.462 | 31.265 | 121.46 | 31.274 |
| 13210 | 121.517 | 31.185 | 121.49 | 31.184 |
| 9204 | 121.449 | 31.313 | 121.486 | 31.342 |
| 6961 | 121.385 | 31.237 | 121.309 | 31.208 |
| 7110 | 121.474 | 31.171 | 121.47 | 31.18 |
| 10573 | 121.493 | 31.315 | 121.503 | 31.341 |
| 12987 | 121.446 | 31.338 | 121.453 | 31.335 |
| 13706 | 121.393 | 31.195 | 121.392 | 31.199 |

、

Table 6: Sample transformed data of start location and end location in the bike sharing data of Mobike in Shanghai on August 20, 2017

| userid | start_location_x | start_location_y | end_location_x | end_location_y |
|---|---|---|---|---|
| 10080 | 348 | 389 | 357 | 388 |
| 13745 | 432 | 18 | 427 | 184 |
| 5564 | 477 | 267 | 441 | 304 |
| 10915 | 447 | 319 | 444 | 314 |
| 11510 | 429 | 349 | 422 | 35 |
| 5198 | 431 | 277 | 443 | 281 |
| 231 | 402 | 263 | 351 | 286 |
| 10127 | 375 | 247 | 376 | 25 |
| 9571 | 491 | 229 | 48 | 224 |
| 5872 | 443 | 262 | 422 | 266 |
| 15094 | 462 | 265 | 46 | 274 |
| 13210 | 517 | 185 | 49 | 184 |
| 9204 | 449 | 313 | 486 | 342 |
| 6961 | 385 | 237 | 309 | 208 |
| 7110 | 474 | 171 | 47 | 18 |
| 10573 | 493 | 315 | 503 | 341 |
| 12987 | 446 | 338 | 453 | 335 |
| 13706 | 393 | 195 | 392 | 199 |

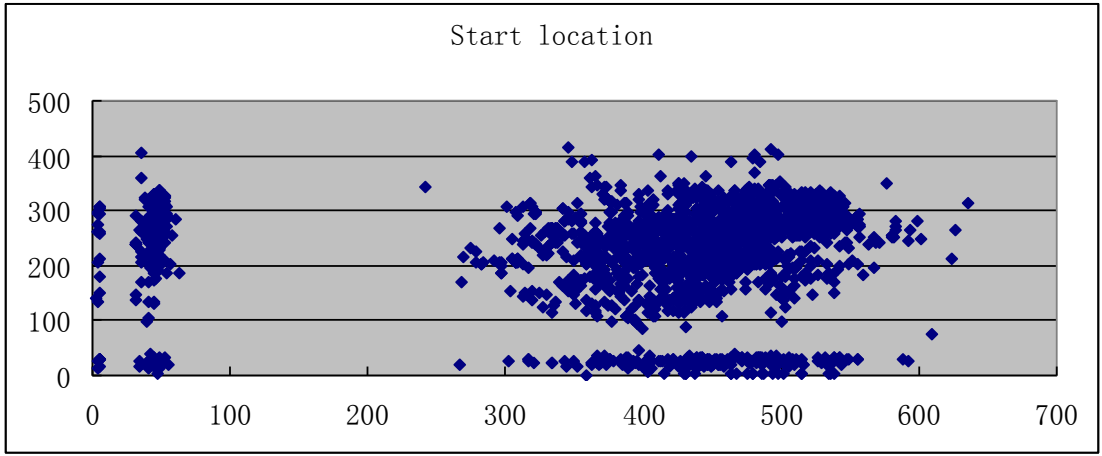

**Figure 6 The scatter plots of the start locations of different users**

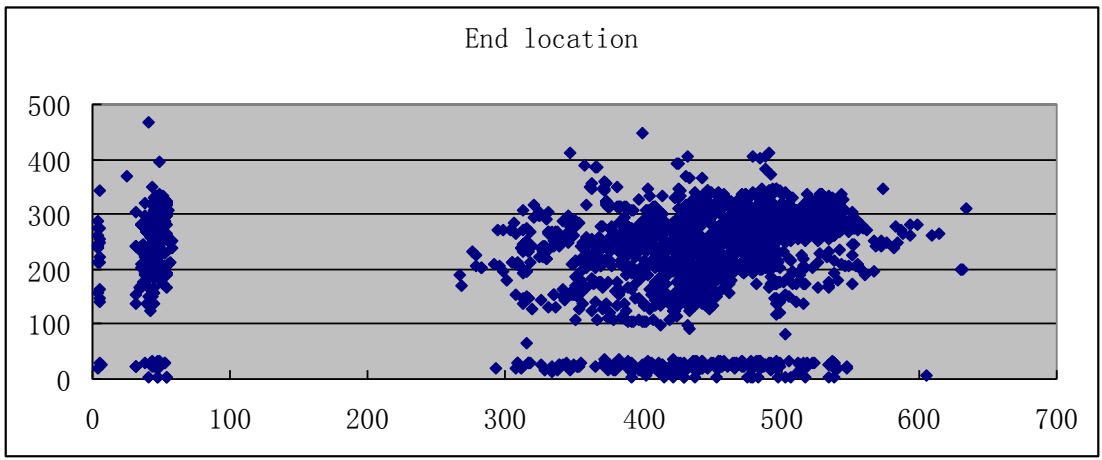

**Figure 7 The scatter plots of the end locations of different users**

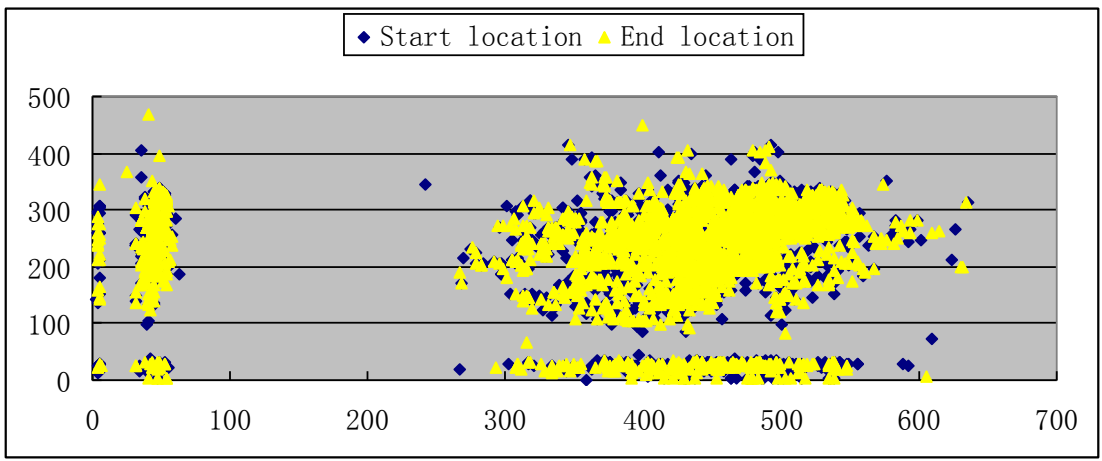

**Figure 8 The scatter plots of the start and end locations of different users**

From the figure 6-8, we can find that the start and end locations are basically overlapping, so that the end locations of some users are the start locations of other users, and the start locations of some users are the end locations of other users, so that sharing services are complementary in space, which fully demonstrates the correctness of the theory we have proposed.

**Mechanism of sharing economy's reconstruction of users consumptions**

In the sharing economy, users often participate fully with multiple roles. Shared users are divided into facility sharing users, platform sharing users, and product sharing users. The shared services platform is built and used by a large number of users themselves. The same user often has the role of service demander and service provider. A large number of users themselves form a closed loop in the shared services platform, which injects the shared services platform with the vitality of the sustainable development. Sharing economic industry is a complete closed-loop industrial chain, while the traditional economic producers and consumers are often independent and separated. For example, in sharing economy, some users ride shared bicycles to provide shared door-to-door services for other users, such as cooking, and picking up children. These users are both providers and consumers of shared services.

The sharing economy can use the complementarity of demand differences for the services corresponding to the resources are shared among users, while traditional economy can not use the complementarity of demand differences for the resources are belong to different users (Figure 9). The shared services platform can schedule resources at any time to provide sharing services for specific users. Since the shared services platform can be infinitely extended through the scheduling of shared services between platforms, it can serve a large number of users. The continuous increase in the number of users will often lead to overwhelming collapse of the traditional platform. However, the increasing number of users will not only exert pressure on the shared services platform, but will further exert the advantages of the shared services platform. The more users, the more balanced the shared services platform by making use of time difference and spatial difference of different users using shared services, and different temporal and spatial resources can be more fully utilized. The non-sharing economy provides users with exclusive resources, while the sharing economy allocates shared resources based on the needs of users. In the non-sharing economy mode, users must purchase resources when they need resources, and resources of different users cannot be shared with each other because there is no sharing mechanism. In the sharing economy model, users do not need to purchase resources when they need resources. They only need to purchase shared services on demand, and the amount, duration, and type of resources that they need to use can be customized when applying for sharing services. When using shared services, There is no need for users to build, upgrade, and maintain service resources because these tasks are already completed within the shared services platform.

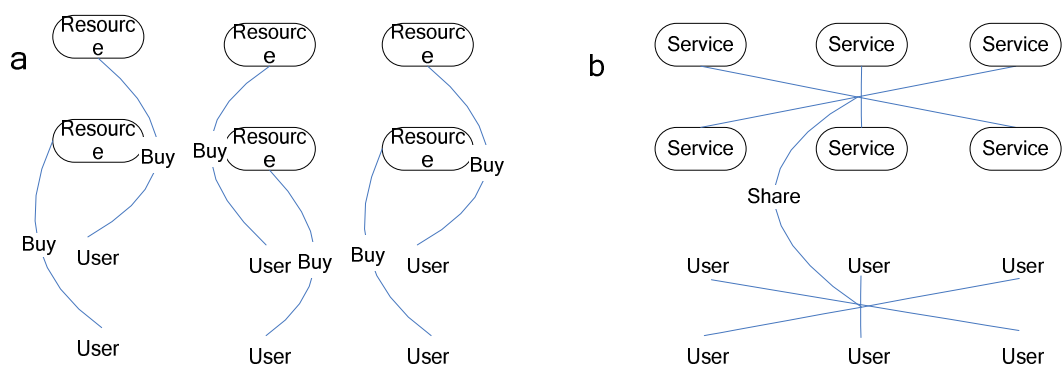

**Figure 9 Demand differences of different users. (a) traditional economy can not use the complementarity of demand differences for the resources are belong to different users, (b) sharing economy can use the complementarity of demand differences for the services corresponding to the resources are shared among users.**

We selected the bike sharing data of Mobike in Shanghai on August, 2017. In the data, the start times of the ride are within August, 2017, as shown in table 7. We retain the orderid , bikeid, userid and start time fields in order to study the relationship between users, resources and services. We counted the number of different orders, different bicycles, and different users in August. At the same time, we randomly selected three days 19, 20, and 21 in August, and performed the same statistics to try to find some rules, as shown in table 8.

**Table 7:Sample data for different orders, different bicycles, and different users**

| orderid | bikeid | userid | start_time |
|---|---|---|---|
| 78387 | 158357 | 10080 | 2016-8-20 6:57 |
| 891333 | 92776 | 6605 | 2016-8-29 19:09 |
| 1106623 | 152045 | 8876 | 2016-8-13 16:17 |
| 1389484 | 196259 | 10648 | 2016-8-23 21:34 |
| 188537 | 78208 | 11735 | 2016-8-16 7:32 |
| 537030 | 66346 | 10335 | 2016-8-7 21:00 |
| 517706 | 99631 | 11258 | 2016-8-29 13:39 |
| 270836 | 63136 | 11361 | 2016-8-29 9:21 |
| 441362 | 331921 | 11705 | 2016-8-31 8:16 |
| 76435 | 347335 | 8135 | 2016-8-30 12:49 |
| 903987 | 13745 | 3117 | 2016-8-9 19:51 |
| 1749105 | 250190 | 8174 | 2016-8-26 17:57 |
| 760441 | 274835 | 17177 | 2016-8-31 9:53 |
| 163153 | 129925 | 9763 | 2016-8-9 8:33 |
| 1718308 | 291555 | 9867 | 2016-8-24 16:32 |

**Table 8: Count the amount of orders, bicycles, and users**

| Date | 2016-8-19 | 2016-8-20 | 2016-8-21 | Aug-16 |
|---|---|---|---|---|
| Orderid | 2425 | 2301 | 1939 | 65535 |
| Bikeid | 2397 | 2264 | 1920 | 54939 |
| Userid | 2167 | 2055 | 1787 | 16132 |

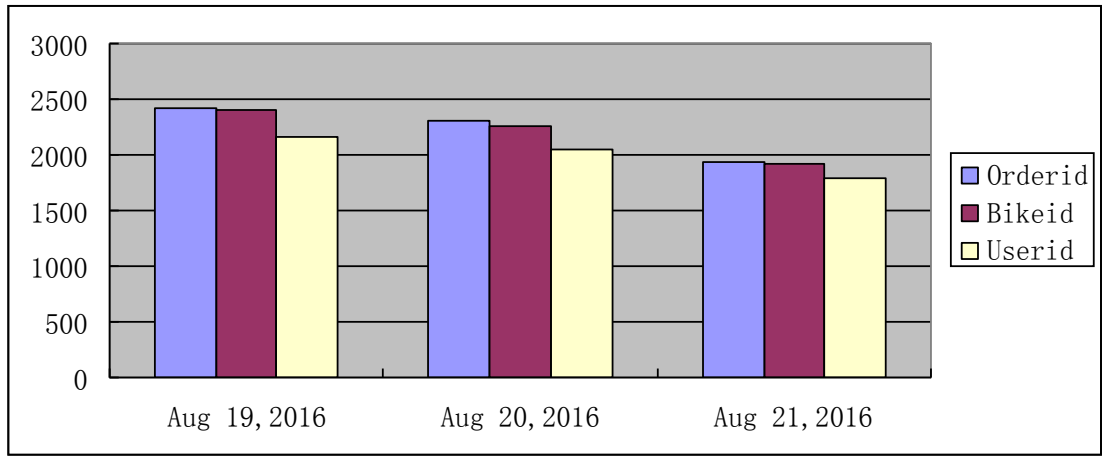

**Figure 10 Histogram of the amount of orders, bicycles, and users.**

As can be seen from Table 8 and Figure 10, the number of orders is greater than the number of users, which means that users use more than one bicycle in the time period, and the number of bicycles is more than the number of users, which further indicates that some users have not only used one bicycles. When bicycles are not shared among users, each user owns a bicycle, so the number of users is equal to the number of bicycles, and our data shows that the inequality between the number of users and the number of bicycles, which can show the sharing feature between them. From the data, we can also see the current problems in the bicycle sharing industry, that is, bicycles are not fully shared among users, and the utilization rate is not very high, because the number of bicycles is more than the number of users. The cost is higher than when they are not shared, which violates the original intention of sharing. Although in the initial stage of the sharing industry, its cost is high, but because of the sharing of services, the convenience to users far exceeds that of non-shared services. For example, if each user can only use his own bicycle, then this user first rides a bicycle to the subway exit near home, however his bicycle can't be taken on the subway, he had to drop his bike and takes the subway to the subway exit near the company, and then the user can not rides his bicycle from the subway exit near the company to the company. This awkward situation is not a problem for shared bikes, because when he reaches the subway near the company, he can ride another shared bike to the company. However, the lack of fully sharing of resources to reduce the cost of shared bicycles will inevitably lead to the dilemma of the shared industry and need to rely on financing to cope with its high costs. This is also the reason why mobike bicycles are closing down and acquired by Meituan.

**Mechanism of sharing economy's reconstruction of industry**

The sharing economy is vertically divided into three major industries and horizontally divided into subdivided industries (Figure 11). The first industry provides facilities as shared services, the second industry provides platforms as shared services, and the third industry provides products as shared services. Platform sharing services can be built on traditional facilities or on shared facilities, but the latter has higher cost performance; product sharing services can be built on traditional platforms or on shared platforms, but the latter has higher cost performance. It can be seen that the third industry is based on the secondary industry and the secondary industry is based on the primary industry. If high-end shared industries are not built on the basis of low-end shared industries, but are built on the basis of traditional industries, the cost of construction will increase, and the effect of construction will decline, because shared industries can provide more personalization services and higher cost performance than traditional industries. After the sharing economy forms a vertical industrial chain, each shared industry sector will benefit from each other. Each shared industrial sector can obtain resources from the neighboring shared industry at the lowest cost and obtain maximum benefits through the provision of shared services, thus creating a win-win situation for the shared industry. So far, the sharing economy model still stays in the sharing of products, lacking sharing of platforms and sharing of facilities. For example, sharing bicycles is a product sharing, which is what we are familiar with; the sharing bicycle production platform is a kind of platform sharing, which has not appeared in our reality; the sharing bicycle production workshop is a kind of sharing of facilities, which has not appeared in our reality. Although there are no existing examples of facility sharing and platform sharing, sharing of platforms and sharing of facilities will surely emerge in the future, thus forming a complete shared industrial chain. All industries in the traditional economic model have a place in the sharing economy or can develop a place in the sharing economy in the future. Sharing bikes, sharing cars, sharing treasures, sharing umbrellas, sharing housing, sharing washing machines, sharing basketball, etc., covers almost all areas that cover clothing, food, housing, and transportation. In the future, there will be more and more things that can be shared, because as long as there are sharing needs, there will be a sharing market. Although the sharing economy continues to subdivide the industry horizontally, each subdivided industry is still vertically divided into three levels: products, platforms, and facilities. For example, traffic sharing services can be divided into traffic product sharing services and traffic platform sharing services, traffic facility sharing services.

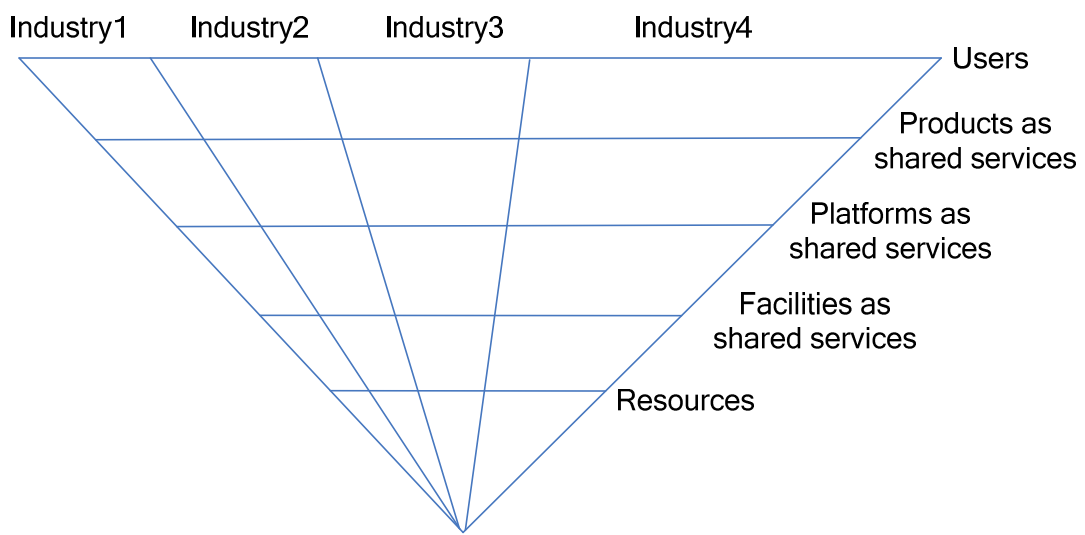

**Figure 11 Three major industries and their subdivided industries in sharing economy.**

Because the information on the Internet is increasing every day, we have count the information that is publicly available online until January 11, 2020. The amount of the related informations about sharing product, sharing platform and sharing facility are counted from the informations in Internet, as shown in table 9.

**Table 9: Amount of informations about sharing product, sharing platform and sharing facility**

| Sharing product | 5,860,000,000 |
| Sharing platform | 1,080,000,000 |
| Sharing facility | 345,000,000 |

From the data in the table 9, it can be seen that the sharing products are currently the most concerned. The sharing platform has gradually received people's attention, and the sharing facilities have received a small amount of attention. These data justify our proposed model.

Shared products, shared platforms and shared facilities constitute a shared industrial chain in the sharing economy, and the suppliers of high level industry are also the users of low level industry in the chain (Figure 12). When users who share products use shared product services, they need to pay the product sharing service fees to the suppliers of the shared products. When users sharing products think that the shared products cannot meet their needs, users sharing the products will require the suppliers of the shared products to upgrade or expand the shared products. In order to upgrade or expand product sharing services, suppliers of shared products will use some of these service fees to upgrade shared products. When the shared product suppliers upgrade or expand the shared products based on the shared platform, at this time the suppliers of shared products become the users of the shared platform. When users of the sharing platforms use the shared platform services, they need to pay platform sharing service fees to the suppliers of the sharing platforms. When the users of the shared platform believe that the shared platforms cannot meet their needs, the users of the shared platform will require the suppliers of the shared platform to upgrade or expand the shared platforms. In order to upgrade or expand the platform sharing services, the suppliers of the sharing platforms will use some of the service fees to upgrade shared platforms. When the suppliers of the shared platform upgrade the shared platform through the shared facilities, the suppliers of the shared platforms become the users of the shared facilities. When users of shared facilities use shared facilities services, they need to pay facility sharing service fees to the suppliers of the shared facilities. When the users of the shared facilities think that the shared facilities cannot meet their needs, the shared facilities users will require suppliers of shared facilities to upgrade or expand shared facilities. In order to expand facilities sharing services, suppliers of shared facilities will use some of the service fees to upgrade or expand shared facilities.

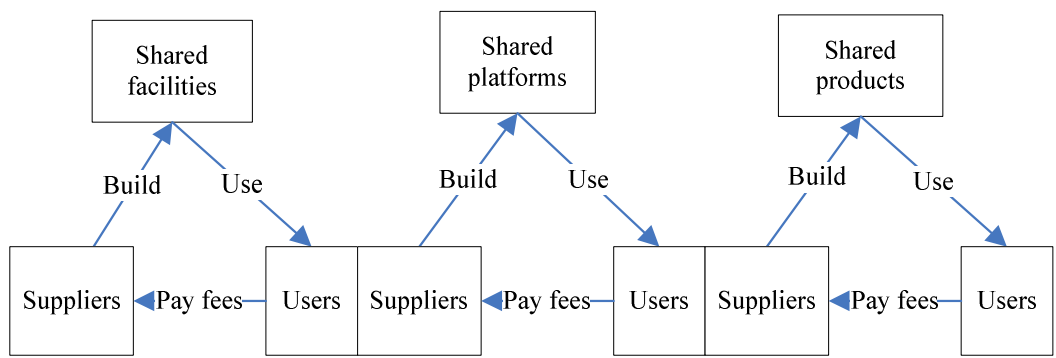

**Figure 12 Shared industrial chain in sharing economy. Shared products, shared platforms and shared facilities constitute a shared industrial chain in the sharing economy, and the suppliers of high level industry are also the users of low level industry in the chain.**

**Mechanism of sharing economy's self-reconstruction in the future**

The sharing economy industry will inevitably move toward two poles: popularity and specialization. On the one hand, the goal of the sharing economy industry is to provide shared services for all users. Therefore, it is necessary to meet the needs of various personalized sharing services among people from all walks of life. This will require the popularity of shared services. At the same time, with the subdivision and development of the sharing economy industry, users in the subdivision industry will inevitably impose higher and higher requirements on the quality of the shared services. This will inevitably require the specialized operation of the sharing service platform by the corresponding subdivided industries, to meet the professional needs of users.

For example, OFO has specialized earlier than other shared bicycle suppliers. OFO uses artificial intelligence, motion detection, Bluetooth unlocking, Internet of Things, GPS/Beidou dual positioning, electronic fences and other technologies to provide people with professional bicycle travel services. The sharing industry for survival of the fittest has now entered an accelerated period of shuffling. Only by carrying out meticulous operations and providing professional services can the users win in the competition in the shared market. Although OFO is also on the verge of collapse, but compared with other shared bicycle companies already adhere to the final. Because the cost of bike sharing is too high, once the investor's money is burned out, the company will be difficult to sustain.

The sharing economy model and other economy models will coexist for a long time, because other economy models are inseparable from the sharing economy model. On the one hand, the sharing economy model is based on other economy models, while other economy models can be formed through the sharing economy model. On the other hand, the sharing economy model may be a local representation of other economy models, and other economy models may also be local representations of the sharing economy model. For example, in the bicycle industry, personal bicycles and shared bicycles coexist, and neither of them will completely replace each other. This coexistence will not affect the development of both. The sharing economy model is to a greater extent a mechanism for resource integration and scheduling, and these resources themselves are still the product of traditional economy models. The sharing economy model can organically combine resources and technologies in the traditional economy model, turn the resources and technologies into shared services in a virtualized manner, and transparently schedule the shared services to users so that users can enjoy convenient and efficient sharing services without needing to know the details of resources and technologies.

Big sharing is the future trend of sharing economy. The superiority of the sharing economy is achieved through the integration and sharing of resources and users. If the shared services in each sharing service platform are isolated from one another, it will be difficult to form a larger and more mature sharing service platform. Of course, companies may establish private shared services for their own internal sharing needs. Private shared services are not connected to public shared services, and seem to be an isolated island. No matter what kind of situation, it will be very difficult for the scale of the sharing service platform to extend.

For example, although mobike has been acquired by Meituan, it has already taken a step forward in big sharing. Mobike's cooperation with Shouqibus prior to its acquisition by the Meituan provided Chinese users with more extensive travel options. Each of the Shouqibus and Mobike has a wealth of industry resources. This cooperation is a win-win situation for both parties. The benefit of this cooperation to Shouqibus is that Mobike brings a large number of user entrances to Shouqibus; the benefit of this cooperation to Mobike is to meets the needs of user scenarios by uniting the shared services of Shouqibus.

# Discussions

We modeled the spatial-temporal differences of different users and the complementarity of spatial-temporal differences in traditional economy and sharing economy to illustrate the sharing economy's reconstructions of time and space. We modeled the demand differences of different users and the complementarity of demand differences in traditional economy and sharing economy to illustrate the sharing economy's reconstructions of users. We modeled the three major industries, their subdivided industries and the shared industrial chain in sharing economy to illustrate the sharing economy's reconstructions of industries.

We compared sharing economy with traditional economy and got their different effects on society, users, industries, future, and thus we conclude that sharing economy can reconstruct these aspects formed in the traditional economy.

Sharing Economy is reconstructing our society. The sharing economy has swept China in recent years and is also sweeping the world，and has become an important technological revolution and business model for the sharing economy can solve a series of problems in the traditional economy. Sharing economy is reconstructing our time and space. The shared services platform can utilize the complementarity of different temporal and spatial resources of different users to balance and coordinate the resource sharing among different users, and thus to increase resource utilization and user satisfaction without adding new resources. Sharing economy is reconstructing our users. In the sharing economy, users often participate fully with multiple roles in a complete closed-loop industrial chain, and the complementarity of demand differences of users can be fully used by the sharing economy. Sharing economy is reconstructing our industry. The sharing economy is vertically divided into three major industries and horizontally divided into subdivided industries, which constitute a shared industrial chain. Sharing economy will self-reconstruct in the future. Firstly, the sharing economy industry will inevitably move toward two poles: popularity and specialization. Secondly, the sharing economy model and other economy models will coexist for a long time. Thirdly, big sharing is the future trend of sharing economy.

**Disclaimer**

The authors declare no competing financial interests.

**Acknowledgements**

The authors acknowledge National New Engineering Research and Practice Project of China (Grant: Higher Education Ministry Letter [2018]17 ); Major Project of National Social Science Fund of China (Grant: 14ZDB101) ; The State Key Program of National Natural Science Foundation of China (Grant: 41630635); National Natural Science Foundation of China(Grant: 61105133); Guangdong Province Higher Education Teaching Research and Reform Project (Grand: Guangdong Higher Education Letter [2016]236) ;Guangdong Province Graduate Education Innovation Project (Grand: 2016JGXM_ZD_30) ;Guangdong Province Joint Training of Graduate Demonstration Base (Grand: Guangdong Teaching and Research Letter [2016]39 ) ; Guangdong Province New Engineering Research and Practice Project (Grand: Guangdong Higher Education Letter [2017]118 ) .

224-233(2019).

[22]Wieland, H., Hartmann, N. N., & Vargo, S. L.. Business models as service strategy. Journal of the Academy of Marketing Science, 45(6), 925-943(2017).